\documentclass[12pt]{article}

\usepackage{amsmath}
\usepackage{amssymb}
\usepackage{mathrsfs}
\usepackage{epsfig}
\usepackage{latexsym}
\usepackage{amssymb}
\usepackage{amsmath}
\usepackage{epsfig}
\usepackage{natbib}
\usepackage[dvips]{color}

\newcommand{\p}{\ensuremath{\mathbb {{}^\prime}}}
\newcommand{\cp}{\ensuremath{\mathbb {CP}}}
\newcommand{\C}{\ensuremath{\mathbb {C}}}
\newcommand{\Z}{\ensuremath{\mathbb {Z}}}
\newcommand{\beq}{\begin{equation}}
\newcommand{\eeq}{\end{equation}}

\def\half{\frac{1}{2}}
\def\halfI{\frac{i}{2}}
\def\Tr{\textrm{Tr}}
\def\H{\textrm{H}}
\def\hsp{,\hspace{.7cm}}

\begin{document}

\begin{titlepage}

\setcounter{page}{1} \baselineskip=15.5pt \thispagestyle{empty}

\begin{flushright}
ULB-TH/07-05\\ hep-th/0702041
\end{flushright}
\vfil

\begin{center}
{\LARGE Trivializing and Orbifolding the Conifold's Base}
\end{center}
\bigskip\

\begin{center}

{\large Jarah Evslin\footnote{ jevslin@ulb.ac.be} and Stanislav Kuperstein\footnote{skuperst@ulb.ac.be}}

\end{center}

\begin{center}
\textit{{\it Physique Th\'eorique et Math\'ematique,
Universit\'e Libre de Bruxelles \& International Solvay
Institutes, ULB Campus Plaine C.P. 231, B--1050 Bruxelles,
Belgium}}
\end{center} \vfil

\noindent 
The conifold is a cone over the space $T^{1,1}$, which is known to be topologically $S^2\times S^3$.  The coordinates used in the literature describe a sphere-bundle which can be proven to be topologically trivializable.  We provide an explicit trivialization of this bundle, with simultaneous global coordinates for both spheres.  Using this trivialization we are able to describe the topology of the base of several infinite families of chiral and non-chiral orbifolds of the conifold. We demonstrate that in each case the 2nd Betti number of the base matches the number of independent ranks in the dual quiver gauge theory.

\vfil

\end{titlepage}

\tableofcontents

\pagestyle{headings}

\section{Introduction}

Type IIB string theory compactified on the conifold, which is topologically a cone over $S^2\times S^3$ \cite{CdO}, is interesting.  It is interesting because it provides a simple example of a background which is holographically dual to an $\mathcal{N}=1$ 4-dimensional gauge theory \cite{KW}.  This gauge theory is even more interesting when one wraps D5-branes around the $\mathit{2}$-cycle in the conifold, as the gauge theory becomes non-conformal and cascades \cite{KS}.  Thus it is clearly of interest to have global coordinates for this $\mathit{2}$-cycle.  Steps in the cascade may be described by NS5-branes that sweep out the conifold's $\mathit{3}$-cycle, eating D3-branes and leaving $H$-flux \cite{KPV,mecascade}.  Thus one would like global coordinates for the $\mathit{3}$-cycle.  Oddly, the coordinates generally used for the conifold do not allow one to simultaneously describe both cycles, instead one may see the base $T^{1,1}$ of the conifold as either an $S^2$ bundle over $S^3$ \cite{CdO} or as an $S^3$ bundle over $S^2$ \cite{KW}.  This bundle may then be proven to be trivializable via its characteristic classes \cite{CdO,KW}, however to describe a brane wrapping a cycle in the base one needs an explicit trivialization.

In this note we will describe the following trivialization.  Define the conifold to be the set of degenerate $2\times 2$ complex matrices $W$, and its base $T^{1,1}$ to be the subset which satisfies
\begin{equation}
\label{eq:WW}
\textrm{Tr} \left( W^\dagger W \right)=1.
\end{equation}
The 3-sphere will be identified with the group $SU(2)$, whose points are special unitary matrices $X$.  The 2-sphere will be identified with the projective line whose elements are projective vectors $v_-$ of unit length.  Then the point $(X,v_-)\in S^3\times S^2$ is identified with the point
\begin{equation} 
\label{def}
W=Xv_-v_-^\dagger
\end{equation}
in $T^{1,1}$.

In Subsec.~\ref{constsec} we will argue that this map is well-defined and invertible and that $W$ is indeed an element of $T^{1,1}$.  Then in Subsec.~\ref{orbsec} we will re-express a large family of group actions on $T^{1,1}$ in terms of the new coordinates, and provide a simple criterion for when they are chiral.  In Sec.~\ref{exsezione} we will consider several infinite families of chiral and non-chiral orbifolds of the conifold, and we will see that in these coordinates one can easily determine the topology of the base, and in particular we find the number of possible non-anomalous wrappings of fractional branes.  In some of these examples it will prove convenient to re-express the three-sphere as a circle bundle over a two-sphere, suggesting that our analysis may be extended to orbifolds of $Y^{p,q}$ \cite{Gauntlett:2004yd}.


\section{The construction} 

We will check the identification (\ref{def}) in two steps.  First we will find a one-to-one identification between the matrices $W$ and a pair $(X,Q)$, where $Q$ is a traceless rank two special unitary matrix parameterizing $S^2$.  Then we will provide a one-to-one map between $Q$ and the projective vectors $v_\pm\in\cp^1\cong S^2$.  


\subsection{Trivializing $T^{1,1}$} \label{constsec}

To see that (\ref{def}) is invertible, notice that given $W$ one may reconstruct $X$, and also a new quantity that we will call $Y$, via the decomposition
\begin{equation} \label{xy}
X \equiv  \textrm{Tr}  (W^\dagger)   \sigma_0 +  (W-W^\dagger)
\qquad \textrm{and} \qquad 
Y \equiv i \textrm{Tr} ( W^\dagger)  \sigma_0 - i (W+W^\dagger),
\end{equation}
where $\sigma_0$ is the $2 \times 2$ unit matrix. Clearly we have $W=X+iY$. Moreover, both $X$ and $Y$
are special unitary matrices: $X^\dagger = X^{-1}$, $Y^\dagger = Y^{-1}$ and $\det X=\det Y=1$.
These properties follow\footnote{The identity $\det A =\half \left( \textrm{Tr} A \right)^2 - \half \textrm{Tr}(A^2)$,
where $A$ is an arbitrary $2\times 2$ matrix, is be useful in the derivation. Also note that the rank $1$ matrix $W$ satisfies  $W^2= \left( \textrm{Tr}  W \right) W$ and similarly for $W^\dagger$ .}
directly from the degeneracy of $W$ and (\ref{eq:WW}).  The information not encoded in $X$ is encoded in $Y$.  However $T^{1,1}$ is 5-dimensional whereas $X$ and $Y$ each run over a 3-dimensional space.  Therefore they are related by a single constraint
\begin{equation} \label{Q}
\Tr Q =0 
\qquad \textrm{where} \qquad
Q \equiv X^{\dagger} Y.
\end{equation}
Since $Q$ is a traceless special unitary matrix it defines an $S^2$,
while the matrix $X$ corresponds to an $S^3$.   $X$ and $Q$ are not related by any constraints, therefore $(X,Q)\in S^3\times S^2$ provides a trivialization of $T^{1,1}$.  Given any point $W$ in $T^{1,1}$ one may find $X$ via (\ref{xy}) and $Q$ via (\ref{Q}).  
This establishes the existence of the map from $W$ to $(X,Q)$.  

The inverse map is simply
\begin{equation} \label{deft}
W=\half( X+iY) = \half X(\sigma_0+iQ)
\end{equation}
and so there is a one to one correspondence between the points $(X,Q)$ on $S^3\times S^2$ and the points $W$ on $T^{1,1}$.  This is the first trivialization that we will provide of $T^{1,1}$, but it is not the trivialization (\ref{def}) that we are seeking.  We still need to show that the $W$ constructed in (\ref{deft}) is an element of $T^{1,1}$, that is, it must be degenerate and it must satisfy (\ref{eq:WW}).  We will now argue that this is indeed the case because
$X$ and $Q$ are both special unitary and $Q$ is traceless.
The fact that $Q$ is unitary and traceless implies that its eigenvalues are $+i$ and $-i$:
\begin{equation} \label{v}
Q v_{\pm} =\pm i v_{\pm},
\qquad \textrm{so} \qquad
W v_+ =0
\end{equation}
and hence $W$ is singular as it has a right zero eigenvector
\footnote{The first equation in (\ref{v}) shows also that $Q$ is necessarily anti-hermitian.}. 
One may also show that $W$ has a left zero eigenvector, using the traceless special unitary matrix 
\begin{equation}
\label{eq:tildeQ}
\widetilde{Q} \equiv Y X^{\dagger}. 
\end{equation}
Now $\widetilde{v}_-^\dagger W=0$, where $\widetilde{v}_-$ is an eigenvector of $\widetilde{Q}$ with eigenvalue $-i$.

The trace of $W^\dag W$ is the sum of its eigenvalues.  We have seen that $v_+$ is an eigenvector with eigenvalue zero.  As $W^\dag W$ is Hermitian, its eigenvectors are orthogonal, which in $\C^2$ means that any vector orthogonal to $v_+$ will be an eigenvector.  Any choice of the eigenvector $v_-$ of $Q$ is orthogonal to $v_+$, so it is an eigenvector of $W^\dag W$.  To demonstrate that $W$ satisfies (\ref{eq:WW}) we therefore need only verify that $v_-$ has eigenvalue one.  Using the anti-hermiticity of $Q$ and the unitarity of $X$ one finds indeed that for the $W$ constructed in (\ref{deft})
\beq
W^\dag Wv_-=\frac{1}{4}(\sigma_0+iQ)X^\dag X(\sigma_0+iQ)v_-=\frac{1}{4}(\sigma_0+iQ) \sigma_0 (1+1)v_-=v_-
\eeq
and so the eigenvalue is equal to $1$ and $W$ is a point on $T^{1,1}$.

Summarizing, we have seen that given a special unitary matrix $X$ and a traceless special unitary matrix $Q$ one may construct a degenerate $W$ satisfying the normalization condition (\ref{eq:WW}) using (\ref{deft}), and that given a degenerate matrix $W$ satisfying (\ref{eq:WW}) one may construct a special unitary matrix $X$ using (\ref{xy}) and a traceless special unitary matrix $Q$ using (\ref{Q}).  This proves that $(X,Q)$ provides a homeomorphism between $T^{1,1}$ and $S^3\times S^2$ as claimed.

On the other hand an isomorphism between the coordinate $Q$ and the projective unit vector $v_{\pm}$ is provided by (\ref{v}), which allows one to explicitly determine $Q$ in terms of $v_\pm$ and an arbitrary choice of 
orthonormal vector $v_\mp$:
\begin{equation} \label{qv}
Q=i v_+ v_+^\dagger - i v_- v_-^\dagger.
\end{equation}
Substituting (\ref{qv}) into (\ref{deft}) one recovers our initial claim (\ref{def}).  Alternately $v_+$ may be obtained directly from $W$, by noting that it is the unique projective zero eigenvector.  This establishes the one-to-one correspondence between points $W$ on $T^{1,1}$ and the pairs $(X,v_\pm)\in S^3\times S^2$.

It is sometimes convenient to introduce explicit coordinates on the spheres,
which can be found using the Pauli matrix decomposition:
\begin{equation}
X = x_0 \sigma_0 + i \sum_i x_i \sigma^i 
\qquad \textrm{and} \qquad
Q =  i \sum_i q_i \sigma^i
\end{equation}
where $\sum_{\mu=0}^4 x_\mu^2=1$ and $\sum_{i=1}^3 q_i^2=1$ are consequences of $\det  X=\det Q=1$.

The 3-sphere described by $X$ can be also viewed as the Hopf fibration $\pi:S^3\rightarrow S^2$. 
One choice of projection map is given by
\begin{equation}
\label{Hopf}
\pi: S^3\longrightarrow S^2: X \mapsto  X Q X^\dagger = \widetilde{Q},
\end{equation}
where in the rightmost expression we have used the definition of $\widetilde{Q}$ in (\ref{eq:tildeQ}).
For this choice of projection map the $U(1)$ action on $X$ is $X \to X e^{\phi Q}$.
Moreover, the matrix $\widetilde{Q}$ provides an alternative description of $T^{1,1}$, since
one may use the pair $(Y,\widetilde{Q})$ instead of $(X,Q)$, which describes $W$ via
\begin{equation}
W = \halfI (\sigma_0 + i\widetilde{Q}) Y.
\end{equation}


\subsection{Group actions on $T^{1,1}$}
\label{orbsec}

The trivializations $(X,Q)$ and $(X,v_\pm)$ are convenient for the study of orbifolds of the conifold, which is our motivation.  This is because left and right actions of the matrix $W$, whose orbifolds create several interesting and famous backgrounds, respect this decomposition. 

We will be interested in orbifolds by group actions that act on the matrix $W$ via left and right matrix multiplication
\beq \label{azione}
W\longrightarrow W\p=\mathcal{O}_L W\mathcal{O}_R,
\eeq
where $\mathcal{O}_L$ and $\mathcal{O}_R$ are both unitary.
As the coordinate matrices $X$ and $Q$ have determinant one, they cannot be simply multiplied by the matrices $\mathcal{O}_{L,R}$, whose determinants are unconstrained.  Instead, to define the group action on $X$ and $Q$, we will need to decompose $\mathcal{O}_{L,R}$ into  terms proportional to the identity matrix and two by two special  
unitary matrices $U_{L,R}$:
\beq \label{gdecomp}
\mathcal{O}_L=e^{i\phi_L}U_L\hsp
\mathcal{O}_R=e^{i\phi_R}U_R,
\qquad \textrm{where} \qquad \det U_{L,R}=1.
\eeq
If the group of transformations is compact then the term proportional to the identity will be a pure phase and $\phi$ will be real.  We will restrict our attention to this case.  

Notice that the decomposition (\ref{gdecomp}) does not uniquely define $\phi$ and $U$, instead they are only defined up to the simultaneous transformation
\beq \label{illdef}
\phi\longrightarrow\phi+\pi\hsp U\longrightarrow -U
\eeq
for the left and or right actions.  We will see momentarily that the group action of $X$ and $Q$ is, as it must be, invariant under the transformation (\ref{illdef}).

The action of the transformation (\ref{azione}) on the coordinates $(X,Q)$ is:
\beq \label{xqazione}
X\longrightarrow X\p=U_LXe^{-(\phi_L+\phi_R)Q}U_R\hsp Q\longrightarrow Q\p=U_R^{\dagger}QU_R.
\eeq
To prove that (\ref{xqazione}) leads to (\ref{azione}) note that $Q$ satisfies the relation 
$e^{\phi Q}=\cos \phi \cdot \sigma_0 + \sin \phi \cdot Q$, which, in turn, follows directly 
from the fact that $Q^2=-\sigma_0$.
Notice also that the determinants of $X\p$ and $Q\p$ are both equal to one 
because the determinants of $U_L$ and $U_R$ are equal to one and, as $Q$ is anti-hermitian and traceless, it exponentiates to a special unitary matrix which also has determinant one.  
Finally, the transformation (\ref{xqazione}) is invariant under (\ref{illdef}), as the transformation of the exponential term is $e^{\pi Q}=-\sigma_0$, which cancels the sign flip of $U_{L,R}$.

Some of the actions (\ref{azione}) respect the chiral $\Z_2$ symmetry of the world-volume gauge theory of D3-branes at the tip of the conifold and D5-branes wrapped on $\mathit{2}$-cycles, while others do not and their quotients lead to chiral world-volume gauge theories.
To see which actions preserve the chiral symmetry, we first express the degenerate matrix $W$ as the dyadic product of two vectors:
\beq \label{dyadic}
W=\left(\begin{array}{c}A_1\\A_2\end{array}\right)\left(\begin{array}{cc} B_1 & B_2\end{array}\right).
\eeq
The chiral symmetry interchanges $A_i$'s and $B_i$'s or, more precisely the action is
\begin{equation}
 W \longrightarrow V_L W^\textrm{T} V_R,
\end{equation}
where $V_L$ and $V_R$ are arbitrary $SU(2)$  matrices.  
Plugging this into (\ref{azione}) and using (\ref{gdecomp}) 
we arrive at the conclusion that the orbifold action (\ref{azione}) preserves the chiral symmetry if and only if
the matrices $U_L^\textrm{T}$ and $U_R$ are conjugate, namely
\begin{equation}
\label{chiral}
U_L^\textrm{T} = V U_R V^\dagger  \qquad \textrm{for some} \qquad V \in SU(2). 
\end{equation}
Geometrically the chiral $\Z_2$ symmetry interchanges the two 2-spheres, $Q$ and $\tilde{Q}^\textrm{T}$, which are transformed by $U_R$ and $U_L^\textrm{T}$ respectively.   Therefore (\ref{chiral}) implies that a symmetry is non-chiral if and only if the transformations of the two 2-spheres are similar.
Furthermore, if $U_L$ and $U_R$ are both diagonal, then (\ref{chiral}) implies that they
are equal, up to a reordering of the diagonal entries.


\section{Examples of orbifolds of the conifold} \label{exsezione}


\subsection{A non-singular chiral quotient}

Perhaps the simplest example of a group action in this framework is the $\Z_k$ group action generated by
left multiplication by the matrix $\mathcal{O}_L=\textrm{diag}(\eta_k,\overline{\eta_k})$:
\beq
\label{gen}
W\longrightarrow W\p=\left(
\begin{array}{cc}\eta_k&0\\0&\overline{\eta_k}
\end{array}
\right)W ,
\eeq
where $\eta_k$ is a $k$th root of unity. This quotient was considered in Ref.~\cite{NOY}.  We see that in this case $\phi_L=\phi_R=0$, $U_R=\sigma_0$
and $U_L=\mathcal{O}_L$.
Using (\ref{xqazione}) one finds the action on $X$ and $Q$ to be
\beq
X\longrightarrow X\p=\left(\begin{array}{cc}\eta_k&0\\0&\overline{\eta_k}\end{array}\right) X\hsp Q\longrightarrow Q\p=Q.
\eeq
Therefore this is the free $\Z_k$ action on the $S^3$ coordinate $X$ and it leaves the $S^2$ fixed.  The quotient is topologically just the product of the Lens space $L(k,1)=S^3/\Z_k$ with the original $S^2$:
\beq
\frac{T^{1,1}}{\Z_k}=L(k,1)\times S^2.
\eeq
This space is a continuous manifold.  

The condition (\ref{chiral}) is not satisfied\footnote{See the comment below (\ref{chiral}).}
and so the world-volume gauge theory on a stack of branes at the tip of a cone over this space is chiral.  Of course the choice of gauge theory depends on more than just the topology of the base, however the quotient contains the relevant geometric data as well and so the condition (\ref{chiral}) is robust.

We will now find the corresponding quiver theory using the strategy described in Refs.~\cite{DM,KS2}.  First we observe that the group action of the generator (\ref{gen}) of $\Z_k$ is
\beq
A_1\longrightarrow A_1\p=\eta_k A_1 \hsp A_2\longrightarrow A_2\p=\overline{\eta_k} A_2 \hsp
B_1\longrightarrow B_1              \hsp B_2\longrightarrow B_2
\eeq
as is apparent from (\ref{dyadic}).
Allowing all possible fractional branes corresponds to allowing all possible actions of the $\Z_k$ group action on the Chan-Paton factors.  All representations of $\Z_k$ are reducible into one-dimensional representations characterized by an integer $j$, in which the generator (\ref{gen}) acts by multiplication by $\eta_k^j$.  The two original gauge groups $SU(r_1)$ and $SU(r_2)$ of the original conifold theory are then decomposed into $\Z_k$ representations where the $j$-representation has multiplicities $N_j^{(1)}$ and $N^{(2)}_j$ respectively.

\begin{figure}[t]
\begin{center}
\leavevmode
\epsfxsize 9   cm
\epsffile{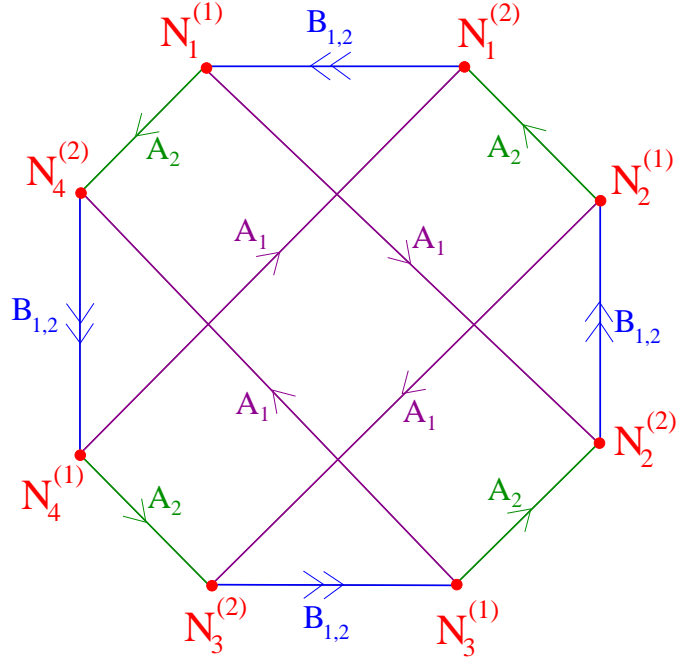}    
\end{center} 
\caption{
This is the quiver diagram for a chiral $\Z_4$ quotient of the conifold which acts on the coordinates $W$ by left multiplication by the matrix $i\sigma_3$.  It consists of 8 gauge groups, but chiral anomaly cancellation demands that their ranks are determined by only two independent parameters, corresponding to the number of D3-branes and the number of D5-branes wrapping the sole non-trivial $\mathit{2}$-cycle in the base $L(4,1)\times S^2$.
\label{faretra} }
\end{figure}

In the parent theory the chiral multiplets $A_i$ and $B_i$ combined to form a non-chiral theory.  Now the generator (\ref{gen}) multiplies $A_1$ by $\eta_k$, and so in the quotient theory $A_1$ will descend to $k$ bi-fundamental chiral multiplets, of which the $j$th connects the $SU(N^{(1)}_j)$ to the $SU(N^{(2)}_{j+1})$.  Similarly $A_2$ is in the conjugate representation so it connects $SU(N^{(1)}_j)$ to $SU(N^{(2)}_{j-1})$.  The $\Z_k$ symmetry group acts trivially on the $B_i$'s, so they both connect the $SU(N^{(2)}_j)$ gauge group to the $SU(N^{(1)}_j)$.  The quiver diagram is then a $2k$-gon whose perimeter is circumnavigated counterclockwise by edges, which are doubled when they extend from an $N^{(2)}$ vertex to an $N^{(1)}$ vertex because there are two $B_i$'s.  In addition there are $A_1$'s which extend clockwise. It can be seen in Fig.~\ref{faretra}. 

Chiral anomaly cancellation at the gauge group $SU(N^{(\alpha +1)}_j)$ demands
\beq \label{anomaly}
2N^{(\alpha)}_j=N^{(\alpha)}_{j+1}+N^{(\alpha)}_{j-1}.
\eeq
Imposing the periodicity condition $N_0=N_k$, the equation (\ref{anomaly}) implies that all $N^{(1)}$'s are equal and also that all $N^{(2)}$'s are equal.  Therefore there are only two independent quantum numbers, the number of D3-branes and the number of D5-branes wrapping a fixed two-cycle.  This corresponds to the fact that the second homology class of the base is rank one by the K\"unneth formula:
\begin{eqnarray}
\H_2(L(k,1)\times S^2)&=&\H_0(L(k,1))\otimes\H_2(S^2)\oplus\H_1(L(k,1))\otimes\H_1(S^2)\\&\oplus&\H_2(L(k,1))\otimes\H_0(S^2)=\Z\otimes\Z\oplus \Z_k\otimes 0\oplus 0\otimes\Z=\Z\nonumber
\end{eqnarray}
and so there is a single $\mathit{2}$-cycle on the base that may be wrapped by a D5-brane.


\subsection{A singular family of chiral and non-chiral orbifolds}

We will now consider a class of orbifolds of the conifold by an action of $\Z_k\times\Z_l$.  The gauge theories of type IIB compactifications on these orbifolds were studied in \cite{Uranga}, and the geometry was investigated in \cite{OT}.  The full quiver was found in the $k=2$, $l=1$ case in \cite{ABFK}.  We will see that, in the $(X,Q)$ coordinates, one can understand the topology of the singular base and its blow up.  In particular we will see that the second Betti number of the blown up base agrees with the number of anomaly-free D5-brane wrappings as calculated in the gauge theory.

The $\Z_k$ action is generated by the transformation
\beq
\label{Wpk}
W\longrightarrow W\p= U_k W U_k
\qquad \textrm{where} \qquad
U_k=\left(
   \begin{array}
    {cc}\eta_{2k}&0\\0&\overline{\eta_{2k}}
   \end{array}
  \right) ,
\eeq
where $\eta_{2k}$ is a $2k$th root of unity.
The action is non-chiral as it satisfies (\ref{chiral}).
In terms of the components of $W$ it reads
\beq
W=\left(\begin{array}{cc}w_1&w_2\\w_3&w_4\end{array}\right)\hsp\hspace{-.2cm}
w_1 \rightarrow \eta_k w_1  \hsp\hspace{-.2cm}  w_4 \rightarrow \overline{\eta_k} w_4 \hsp\hspace{-.2cm} (w_2,w_3)\rightarrow (w_2,w_3).
\eeq
The $\Z_l$ action is also non-chiral and is given by
\beq
\label{Wpl}
W\longrightarrow W\p= U_l W U_l^\dagger
\qquad \textrm{where} \qquad
U_l=\left(
   \begin{array}
    {cc}\eta_{2l}&0\\0&\overline{\eta_{2l}}
   \end{array}
  \right) 
\eeq
or, in terms of the conifold components
\beq
(w_1,w_4)\rightarrow (w_1,w_4) \hsp w_2 \rightarrow \eta_l w_2  \hsp  w_3 \rightarrow \overline{\eta_l} w_3.
\eeq
To avoid writing both actions on $W$, we will restrict our attention to the case in which $k$ and $l$ are relatively prime.  
The product of the cyclic symmetries is now a single cyclic group
\beq
\Z_k\times\Z_l=\Z_{kl},
\eeq
which is generated by a single element
\begin{eqnarray} 
\label{xqazioneb}
X  &\rightarrow&  U_k U_l X  U_l^\dagger U_k = 
   \left(\begin{array}{cc}\eta_{2k}\eta_{2l}&0\\0&\overline{\eta_{2k}}\overline{\eta_{2l}}\end{array}\right)
   X\left(\begin{array}{cc}\eta_{2k}\overline{\eta_{2l}}&0\\0&\overline{\eta_{2k}}\eta_{2l}\end{array}\right)
\nonumber \\
Q  &\rightarrow&  U_k^\dagger U_l Q  U_l^\dagger U_k =
   \left(\begin{array}{cc}\overline{\eta_{2k}}\eta_{2l}&0\\0&\eta_{2k}\overline{\eta_{2l}}\end{array}\right)
   Q\left(\begin{array}{cc}\eta_{2k}\overline{\eta_{2l}}&0\\0&\overline{\eta_{2k}}\eta_{2l}\end{array}\right),
\end{eqnarray}
as one can derive from (\ref{Wpk}), (\ref{Wpl}) and  (\ref{xqazione}).  It is straightforward to see that the $\Z_{kl}$ action does not break the chiral symmetry if and only if 
$k=1$ or $l=1$, when it reduces to (\ref{Wpl}) or (\ref{Wpk}) respectively.

The action (\ref{xqazioneb}) on $Q$ is easy to interpret.  It is the conjugation of special unitary traceless matrix by an exponent of $\sigma_3$, which corresponds to a rotation of the two-sphere $S^2_Q$ parametrized by $Q$ about the $z$ axis.  This means that the action on $S^2_Q$ has two fixed points, the north and south poles.  The angle of the rotation is irrelevant and depends on our representatives of the roots of unity, the important quantity is the order of the rotation, that is the number of times that it must be performed for the sphere to return to its original position.  The order is $kl$.  This does not mean that there will necessarily be a singularity at each pole, as the action acts simultaneously on the 3-sphere parameterized by $X$.  Singularities will occur where both $X$ and $Q$ are fixed.

The action (\ref{xqazioneb}) on the 3-sphere parametrized by $X$ is more difficult to interpret.  It will be useful to decompose the 3-sphere as the Hopf fibration $\pi_P:S^3\rightarrow S^2_P$ over a 2-sphere which we name $S^2_P$ using the projection map
\beq \label{pi}
\pi_P:S^3\longrightarrow S^2_P:X\mapsto P=iX\sigma_3 X^\dag.
\eeq
This choice of projection map differs from (\ref{Hopf}) and it is not canonical, 
since one may use any linear combination of the Pauli $\sigma$ matrices with determinant equal to $-1$. However, the choice of $\pi_P$ in (\ref{pi}) will lead to simplifications because $\sigma_3$ commutes with the group actions, and in particular with $U_L=U_k U_l$.  
This means that the group action on $P$ is just
\beq
P\longrightarrow P\p=U_LPU_L^\dag=\left(\begin{array}{cc}\eta_{2k}\eta_{2l}&0\\0&\overline{\eta_{2k}}\overline{\eta_{2l}}\end{array}\right) P \left(\begin{array}{cc}\overline{\eta_{2k}}\overline{\eta_{2l}}&0\\0&\eta_{2k}\eta_{2l}\end{array}\right)
\eeq
which, like the group action on $Q$, is simply a rotation about the $z$ axis.

The group action on the product of the two 2-spheres $S^2_P\times S^2_Q$ is a rotation about the $z$-axis of each, and so it has four fixed points corresponding to the products of the north and south $P$ and $Q$ poles.  This does not mean that the total action on $T^{1,1}$ has four fixed points, because $T^{1,1}$ is a circle bundle over $S^2_P\times S^2_Q$ and the action is in general non-trivial on the circle fiber.  To find and analyze the singularities one then needs to understand the group action on the circle fiber above the four products of poles.  Note that every fiber is a circle of the form $Xe^{i\phi\sigma_3}$, for some $X$, which is parametrized by $\phi$.  In particular every point on this circle projects to the same point
\beq
\pi_P(X)=\pi_P(Xe^{i\phi\sigma_3})=iX\sigma_3 X^\dag.
\eeq
As the circle $Xe^{i\phi\sigma_3}$ is only non-trivially fibered over $S^2_P$, the circle action will be the same at the north and south poles of $S^2_Q$, and so we need only understand the action on $\phi$ at the poles of $S^2_P$.

\begin{figure}[t]
\begin{center}
\leavevmode
\epsfxsize 14  cm
\epsffile{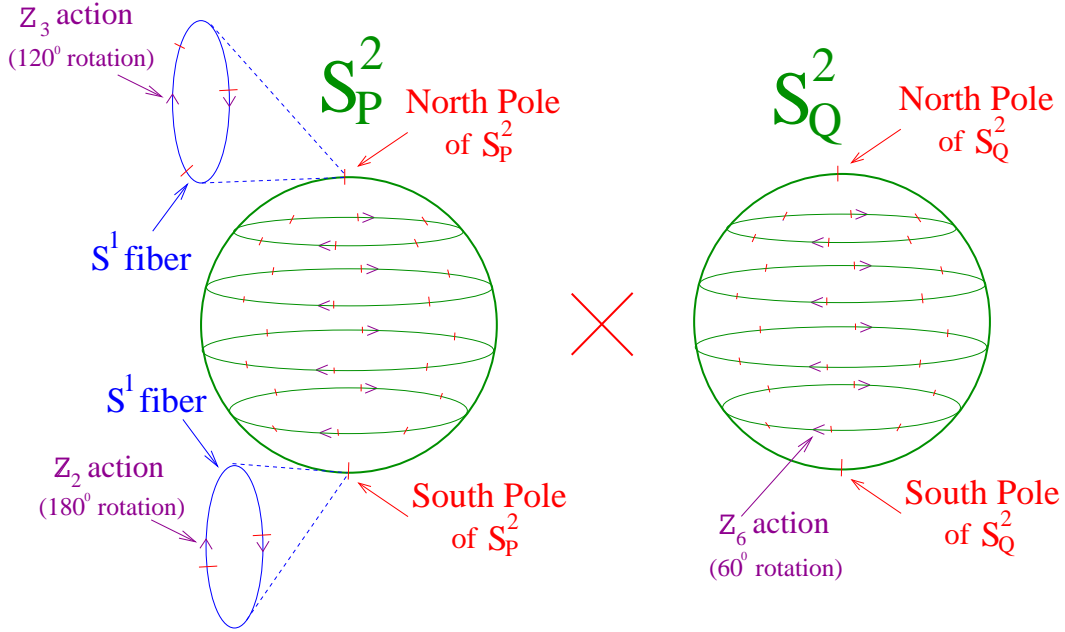}    
\end{center} 
\caption{
This is the action of $g$ on $T^{1,1}$ in the case of the $k=3$, $l=2$, $\Z_6$ chiral orbifold of the conifold, which is a circle bundle over the product of two-spheres $S^2_P\times S^2_Q$.  It rotates each two-sphere by 60 degrees, leaving four fixed points at the products of the poles.  At these fixed points the group $\Z_6$ does not act freely on the circle fiber.  Instead, at the two fixed points corresponding to the north pole of $S^2_P$ only a $\Z_3$ subgroup acts freely, while at the other two fixed points, which correspond to the south pole of $S^2_Q$, only a $\Z_2$ subgroup acts freely.
\label{sfere} }
\end{figure}

We will define the north pole of $S^2_P$ to be the point
\beq
P=i\sigma_3,
\eeq
where $X=e^{i\phi\sigma_3}$ is the circle fibered over the north pole in $S^3$.  Using (\ref{xqazioneb}) one can now find the group action on $X$ at the circle above the north pole.  If for concreteness we set $\eta_p=e^{2\pi i/p}$ then
\begin{eqnarray} \label{santa}
X=e^{i\phi\sigma_3}  \rightarrow  X \p &=&
\left(\begin{array}{cc}\eta_{2k}\eta_{2l}&0\\0&\overline{\eta_{2k}}\overline{\eta_{2l}}\end{array}\right)e^{i\phi\sigma_3}\left(\begin{array}{cc}\eta_{2k}\overline{\eta_{2l}}&0\\0&\overline{\eta_{2k}}\eta_{2l}\end{array}\right)\nonumber\\&=&\left(\begin{array}{cc}e^{i(\phi+2\pi/k)}&0\\0&e^{-i(\phi+2\pi/k)}\end{array}\right)
\end{eqnarray}
and so we see that the action on $\phi$ is
\beq \label{phiazione}
\phi\longrightarrow\phi+2\pi/k.
\eeq
Therefore the generator $g$ of $\Z_{kl}$ does not fix any point $\phi$ in the fiber over the north pole of $S^2_P$ 
and at either pole of $S^2_Q$, instead it translates $\phi$ around the circle fiber by $2\pi/k$ as in (\ref{phiazione}).  However, if $l\neq 1$ then the element $g^k$ is a non-trivial element of the group $\Z_{kl}$ and it does fix the position on the fiber of every point $\phi$.  The element $g^k$ is order $l$, and so we find that the entire fiber is a $A_{l-1}$ singularity.  In fact, there are two circles of $A_{l-1}$ singularity, one at the north pole of $P$ and the north pole of $Q$ and one at the north pole of $P$ and the south pole of $Q$.  These two circles of $A_{l-1}$ singularity were found in \cite{Uranga}.

Similarly one may compute the action of the group on the circle fibered over the south pole of $S^2_P$, corresponding to the point
\beq
P=-i\sigma_3\hsp X=i\sigma_1e^{i\phi\sigma_3}.
\eeq
Now $X$ anti-commutes with $\sigma_3$, and so the $\eta_k$'s from the right and left action cancel, instead of the $\eta_l's$ as in the north pole case (\ref{santa}).  This means that $g$ translates $\phi$ by $2\pi/l$ and so one finds a circle of $A_{k-1}$ singularity fibered over the product of the south pole of $S^2_P$ with the north pole of $S^2_Q$ and another fibered over the product of the south pole of $S^2_P$ with the south pole of $S^2_Q$.  These actions are schematically displayed in Fig.~\ref{sfere}.

In all, the quotient $T^{1,1}/Z_{kl}$ contains 2 circles of $A_{k-1}$ singularity and 2 circles of $A_{l-1}$ singularity, as found in \cite{Uranga}.  Blowing up these singularities until the base becomes smooth one expects to find $l-1$ $\mathit{2}$-cycles above the product of the two north poles, $l-1$ $\mathit{2}$-cycles above the $P$-north pole times the $Q$-south pole, $k-1$ $\mathit{2}$-cycles above the $P$-south pole times the $Q$-north pole and $k-1$ $\mathit{2}$-cycles above the $P$-south pole times the $Q$-south pole.  
The $\mathit{2}$-cycles at $Q$-north and $Q$-south poles are not homologous. To see this, recall that according to the McKay correspondence the intersection matrix in the 4-dimensional base of the $\mathit{2}$-cycles at one pole is the weight lattice of $A_n$ 
(in our case $n=k-1$ and $n=l-1$).   In the full 5-dimensional space one instead must consider intersection numbers of $\mathit{2}$-cycles and $\mathit{3}$-cycles.  Notice that each $\mathit{2}$-cycle in the 
blowup corresponds to a unique $\mathit{3}$-cycle which is its orbit with respect to the circle action.  
The intersection numbers in the 5-dimensional space of the $\mathit{2}$-cycles with the corresponding 
$\mathit{3}$-cycles are again given by the weight lattice of $A_n$.
On the other hand, as they are separate in $Q$, the intersection number of any $Q$-northern $\mathit{2}$-cycle with 
any $\mathit{3}$-cycle which is the product of a $Q$-southern $\mathit{2}$-cycle with the circle is zero.  
As the weights of $A_n$ are linearly independent, no linear combination is zero and so no northern cycle
is a linear combination of southern cycles and vice versa.

To summarize, one expects $2(k+l)-4$ different $\mathit{2}$-cycles to be created by the blow-up.  There was already one $\mathit{2}$-cycle in $T^{1,1}$, and so the total number of $\mathit{2}$-cycles is expected to be\footnote{ 
This result follows also from the toric diagram of the orbifold, which is given by a $k \times l$ box (see \cite{Uranga}).}
\beq
N=2(k+l)-3. \label{N}
\eeq
We will now compare the result for $N$ in (\ref{N}) with the number of independent gauge groups in the corresponding quiver theories.

To find the corresponding quiver theories, we again decompose $W$ into a dyadic product of two vectors as in (\ref{dyadic}).  An action of $g$ on the components of the factors is easily found, as $\mathcal{O}_L$ multiplies $A$ from the left and $\mathcal{O}_R$ multiplies $B$ from the right.  However the vectors $A$ and $B$ are not entirely determined by $W$, they may be rotated in opposite directions.  We will use this freedom to define the simpler action
\beq \label{aazione}
A_1\rightarrow \eta_k\eta_l A_1\hsp A_2\rightarrow A_2\hsp
B_1\rightarrow \overline{\eta_l} B_1\hsp B_2\rightarrow \overline{\eta_k}B_2.
\eeq

\begin{figure}[t]
\begin{center}
\leavevmode
\epsfxsize 9   cm
\epsffile{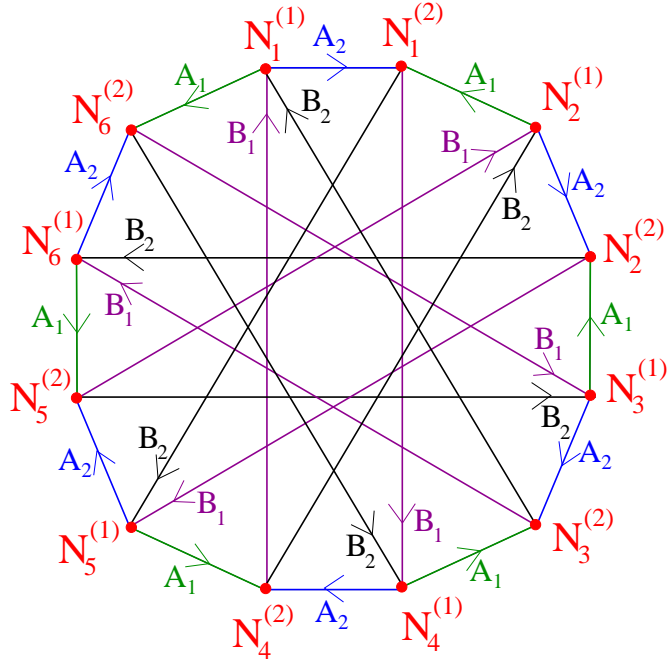}    
\end{center} 
\caption{ 
This is the quiver diagram for $k=3$, $l=2$, $\Z_6$ chiral orbifold of the conifold.  It consists of 12 gauge groups, but chiral anomaly cancellation demands that their ranks are determined by only 8 independent parameters, indicating the presence of $2k+2l-3=7$ two-cycles on which fractional branes can be wrapped.
\label{klfaretra} }
\end{figure}

As in the previous example, we decompose the original gauge group $SU(r_1)\times SU(r_2)$ into one dimensional representations of the orbifold group $\Z_{kl}$.  We let $N^{(\alpha=1,2)}_j$ denote the multiplicity of the representation $\rho(g)=\eta_{lk}^j$ in the parent gauge group $SU(r_\alpha)$.  A node in the quiver diagram corresponds to one of these representations, and so the quiver diagram will have $2kl$ nodes.  The action of $g$ on $A_1$ in (\ref{aazione}) is the element $k+l\in\Z_{kl}$ and so the $kl$ chiral multiplets descending from $A_1$ extend from the node $N^{(1)}_j$ to the node $N^{(2)}_{j+k+l}$.  Similarly the action on $A_2$ is the identity element $0$ and so the corresponding arrows extend from $N^{(1)}_j$ to $N^{(2)}_j$.  The action on $B_1$ is the element $-k$ and so the $B_1$ arrows extend from $N^{(2)}_j$ to $N^{(1)}_{j-k}$.  Finally the $B_2$ arrows extend from $N^{(2)}_j$ to $N^{(1)}_{j-l}$.  This quiver is illustrated in Fig.~\ref{klfaretra}.  Notice that if $l=1$ then there are only $k$ components in each group and so the above indices should be read modulo $k$, which implies that $A_1$ arrows run anti-parallel to $B_2$ arrows and $A_2$ arrows to $B_1$, therefore the theory is non-chiral as we have seen.  Similarly if $k=1$ then $A_1$ and $B_1$ are anti-parallel as are $A_2$ and $B_2$.

The chiral anomaly cancellation condition, that the same number of arrows enter each mode as leave, is that
\beq \label{nrel}
N^{(\alpha)}_j+N^{(\alpha)}_{j+k+l}=N^{(\alpha)}_{j+k}+N^{(\alpha)}_{j+l},
\eeq
where $\alpha=1,2$ and the subscripts are understood to be modulo $kl$.
Of these $kl$ conditions on $kl$ ranks for each value of $\alpha$, 
only $kl-k-l+1$ are linearly independent. This leaves $k+l-1$ free ranks in each parent gauge group, for a total of $2k+2l-2$ choices of ranks.  Notice, for example that if you know $k+l$ consecutive values of $N^{(\alpha)}_j$ at fixed $\alpha$ then you can use (\ref{nrel}) to extend this set arbitrarily far in any direction, but there will be a single constraint which enforces periodicity, leaving $k+l-1$ free parameters at each $\alpha$.  The overall rank corresponds to the D3-brane charge, leaving $2k+2l-3$ choices determined by the consistent D5-brane wrappings on $\mathit{2}$-cycles, in agreement with (\ref{N}).
We provide a more rigorous proof of this statement in Appendix \ref{App}.

\section{Conclusions}
\label{conc}
In this note we have provided global coordinates for the $S^2$ and $S^3$ whose Cartesian product is diffeomorphic to $T^{1,1}$, the base of the conifold.  We have seen that several classes of orbifolds of the conifold are easily described in these coordinates, and we have used them to find the topology of a one-parameter infinite family of chiral orbifolds, and the corresponding quiver gauge theories.  We have calculated the second Betti numbers of several families of orbifolds of the conifold and have compared them successfully with gauge theory expectations.

It would be of interest to extend this decomposition to $Y^{p,q}$ and $L^{a,b,c}$ spaces, as the $S^3$ in these cases is swept out by a generalization of the NS5-brane in \cite{KPV}, as will be elaborated in \cite{EKK}.


\appendix
\section{Counting independent ranks}

\label{App}
In this appendix we will demonstrate that the adjacency matrix defined by the relation (\ref{nrel})
has $2(k+l)-2$ independent zero eigenvectors. 
One of these eigenvectors assigns identical rank to all the gauge groups, so it describes physical D3-branes.
The remaining $2(k+l)-3$ zero eigenvectors correspond to fractional D3-branes which are D5-branes wrapping the $\mathit{2}$-cycles that we have found 
in the last section.

First, an arrow which extends $k$ spaces clockwise acts on the $N_j$ via $U^{j}$, where  $U$  is the $kl \times kl$ shift matrix
\begin{equation}
U = \left(
    \begin{array}{cccccc}
      0      & 1        & 0       & 0       & 0       & \ldots  \\
      0      & 0        & 1       & 0       & 0       & \ldots  \\
      \vdots & \vdots   & \vdots  & \ddots  &         &         \\
      0      & 0        & 0       & \ldots  & 0       & 1       \\
      1      & 0        & 0       & \ldots  & 0       & 0
    \end{array}
\right).
\end{equation}
Now the $2kl \times 2kl$ adjacency matrix $C$ corresponding to (\ref{nrel}) is
\begin{equation}
C = \left(
    \begin{array}{cc}
      0      & D       \\
      -D^\textrm{T}      & 0        
    \end{array}
\right),
\qquad 
\textrm{where}
\qquad 
D \equiv \mathbf{1} + U^{k+l} - U^k -U^l.
\end{equation}
Furthermore, a zero eigenvector of $C$ is may be decomposed as $v=(v_1, v_2)$, 
where $v_1$ and $v_2$ are left and right zero eigenvectors
of the matrix $D$. Notice that the eigenvalues of the $kl \times kl$ shift matrix $U$ are $\omega^{i=1,\ldots,kl}$,
where $\omega$ is defined by $\omega^{kl}=1$. Thus for an eigenvector $v_\omega$ corresponding to the eigenvalue $\omega$
we have

\begin{equation}
D v_\omega = \left( 1 + \omega^{k+l} - \omega^k - \omega^l \right) v_\omega = (1-\omega^k)(1-\omega^l) v_\omega.
\end{equation}
We conclude therefore that for $v_w$ to be a zero eigenvector we need $\omega^k=1$  or $\omega^l=1$.
This gives $(k+l-1)$ independent vectors since for the vector $(1,1,\ldots,1)$ both conditions are satisfied simultaneously.
After a discrete Fourier transform one obtains a basis for the zero eigenvectors in which all entries are $0$ or $1$.
This is consistent with the integrality of the ranks of the gauge groups.
Finally, we recall that a zero eigenvector of $C$ is determined by two independent zero eigenvectors of $D$,
so the total multiplicity is $2(k+l)-2$, completing our proof.

\section* {Acknowledgement}

We would be honoured to thank R. Argurio, F. Bigazzi, A. Hanany, C. Krishnan and D.~Persson for invaluable comments.


\bibliographystyle{unsrt}

\end{document}